\documentclass[12pt,a4paper]{article}
\usepackage{cite}
\usepackage{fullpage}
\usepackage[centertags]{amsmath}
\allowdisplaybreaks[4]
\usepackage{amssymb}
\usepackage{bm}
\usepackage[dvips]{graphicx}
\usepackage{url}
\usepackage[hyperindex]{hyperref}
\usepackage{color}
\usepackage{verbatim}

\newcommand{\half}{\frac{1}{2}}
\newcommand{\Vcc}{V_{\mathrm{CC}}}
\newcommand{\Vnc}{V_{\mathrm{NC}}}
\newcommand{\Rnc}{R_{\mathrm{NC}}}
\newcommand{\VR}{V_{\mathrm{R}}}

\newcommand{\dM}{\delta_{\mathrm{M}}}

\newcommand{\averP}[2][e]{\overline{P}^{\mathrm{#2}}_{\nu_{#1} \to \nu_\beta}}

\newcommand{\psiV}{\psi^{\mathrm{V}}}
\newcommand{\PsiV}{\Psi^{\mathrm{V}}}
\newcommand{\psiM}{\psi^{\mathrm{M}}}
\newcommand{\PsiM}{\Psi^{\mathrm{M}}}
\newcommand{\pVm}{
  \begin{pmatrix}
    \psiV_1 \\
    \psiV_2
  \end{pmatrix}}
\newcommand{\PsiVt}{\tilde{\Psi}^{\mathrm{V}}}

\newcommand{\PsiMt}{\tilde{\Psi}^{\mathrm{M}}}
\newcommand{\PsiIt}{\tilde{\Psi}^{\mathrm{I}}}

\newcommand{\HF}{\mathcal{H}^{\mathrm{F}}}
\newcommand{\HV}{\mathcal{H}^{\mathrm{V}}}
\newcommand{\HM}{\mathcal{H}^{\mathrm{M}}}
\newcommand{\HVt}{\tilde{\mathcal{H}}^{\mathrm{V}}}

\newcommand{\HMad}{\HM_{\mathrm{ad}}}
\newcommand{\HMna}{\HM_{\mathrm{na}}}

\newcommand{\SM}{S^{\mathrm{M}}}
\newcommand{\SMpert}{\SM_{\mathrm{pert}}}
\newcommand{\SMpX}{\SMpert(x_f,x_i)}

\newcommand{\SMXad}{\SM_\mathrm{ad}(x_f,x_i)}
\newcommand{\SMXada}{\SM_\mathrm{ad}(x,x_i)}

\newcommand{\Ut}{\tilde{U}_2}

\newcommand{\dx}{\mathrm{d}x}
\newcommand{\ddx}{\frac{\mathrm{d}}{\mathrm{d} x}}
\newcommand{\df}{\dot{\varphi}}
\newcommand{\dw}{\dot{\omega}}
\newcommand{\intx}{\int_{x_i}^{x_f}}
\newcommand{\mIm}{\mathrm{Im}}
\newcommand{\mRe}{\mathrm{Re}}

\newcommand{\hrx}{\hat{\rho}(x)}
\newcommand{\rF}{\rho^{\mathrm{F}}}
\newcommand{\rV}{\rho^{\mathrm{V}}}
\newcommand{\PauliV}{\vec\sigma_{\mathrm{V}}}
\newcommand{\vB}{\vec{B}}
\newcommand{\vS}{\vec{S}}
\begin{document}

\vspace{1cm}
\begin{center}
{\large\bf CP-violating Phases in Active-Sterile Solar Neutrino
Oscillations}
\end{center}
\vspace{0.3cm}

\begin{center}
{\bf H.W. Long} \footnote{E-mail: lhw0128@mail.ustc.edu.cn}\\
Department of Modern Physics, University of Science and\\
Technology of China, Hefei, Anhui 230026, China \\

\vspace{0.5cm}
{\bf Y.F. Li} \footnote{E-mail: liyufeng@ihep.ac.cn}\\
Institute of High Energy Physics, Chinese Academy of\\
Sciences, Beijing 100049, China \\

\vspace{0.5cm}
{\bf C. Giunti} \footnote{E-mail: giunti@to.infn.it}\\
INFN, Sezione di Torino, Via P. Giuria 1, I--10125
Torino, Italy
\end{center}

\setcounter{footnote}{0}

\vspace{1.5cm}
\begin{abstract}
Effects of CP-violating phases in active-sterile solar neutrino
oscillations are discussed in a general scheme of 3+$N_{s}$ mixing,
without any constraint on the mixing between the three active and
the $N_{s}$ sterile neutrinos, assuming only a realistic hierarchy of
neutrino mass-squared differences. A generalized \emph{Parke} formula
describing the neutrino oscillation probabilities inside the Sun is calculated.
The validity of the analytical calculation and the probability
variation due to the unknown CP-violating phases are illustrated
with a numerical calculation of the evolution equation in the case of 3+1 neutrino mixing.

\end{abstract}
\vspace{1.5cm}

\text{PACS numbers:  14.60.St, 13.35.Hb, 14.60.Pq, 26.65.+t}

\newpage

\section{Introduction}
\label{introduction}

Despite the success of standard three-neutrino oscillations
\cite{PDG} in explaining the results of solar (SOL), atmospheric
(ATM), reactor and accelerator neutrino experiments with two distinct
mass-squared differences (i.e., $\Delta{m}^{2}_{\mathrm{SOL}}$ and
$\Delta{m}^{2}_{\mathrm{ATM}}$) and three non-zero mixing angles
\footnote{The mass-squared differences and the three mixing angles are
defined according to the standard parametrization in the latest PDG
publication \cite{PDG}.}(i.e., $\theta_{12}$, $\theta_{23}$ and
$\theta_{13}$), some anomalies in short baseline
(SBL) neutrino oscillation experiments (e.g., LSND \cite{LSND},
MiniBooNE \cite{Mini}, the Reactor anomaly \cite{Reactor} and
Gallium anomaly \cite{Gallium}) indicate the existence of
oscillations with much shorter baselines. This would imply the
existence of extra mass-squared differences (i.e.,
$\Delta{m}^{2}_{\mathrm{SBL}}$) with the hierarchy
\begin{align}
  \Delta{m}^{2}_{\mathrm{SOL}} \ll \Delta{m}^{2}_{\mathrm{ATM}} \ll
  \Delta{m}^{2}_{\mathrm{SBL}}\quad,
\label{hierarchy}
\end{align}
and therefore the mixing of three active neutrinos with extra
sterile neutrino states \cite{Giunti11,Giunti12,Schwetz4}.
Furthermore, the analysis \cite{Raffelt,Ade:2013lta} of cosmic
microwave background and large scale structure data may hint at the
existence of additional radiation in the Universe, with sterile
neutrinos being one of the plausible candidates. One extra sterile
neutrino is also allowed by recent analyses of Big Bang
Nucleosynthesis \cite{Mangano}. Therefore, we should be open-minded
on the existence of sterile neutrinos and it might be instructive to
study the effects of the light sterile neutrino hypothesis in solar
neutrino \cite{solarste,palazzo1,palazzo2} and atmospheric neutrino \cite{ice}
oscillation experiments or in non-oscillation processes including
beta decay \cite{beta} and neutrinoless double-beta decay
\cite{double}.

Solar neutrinos produced in the core of the Sun can undergo
matter-enhanced Mikheev-Smirnov-Wolfenstein (i.e., MSW
\cite{Wolfenstein:1978ue,Mikheev:1985gs}) oscillations when they propagate
inside the Sun. When a generic scheme with one or more
sterile neutrinos is considered, some sub-leading effects including
the effects of light sterile neutrinos \cite{whitepaper}, non-unitarity of the
lepton mixing matrix (NU) \cite{NU} and non-standard interactions
(NSI) \cite{NSI} may contribute to the neutrino oscillation
probabilities and modify the standard MSW picture in the
three-neutrino mixing scheme. Therefore, it is possible to constrain
or measure these high-order effects with future precision solar neutrino
experiments.

In an earlier work \cite{solarste} by two of the present authors, we
proposed a general method to calculate the matter effects of solar
neutrino oscillations with an arbitrary number of sterile neutrinos,
without any constraint on the magnitudes of the active-sterile
mixing. However, the oscillation probabilities derived in Ref.~\cite{solarste} are only valid
for a real neutrino mixing matrix because of an incorrect treatment of the CP-violating phases.
As will be explained in Section~\ref{sec:parke},
in Ref.~\cite{solarste} the effects of the CP-violating phases have been removed with
an inappropriate phase-transformation.

In this work, we study the
evolution of the neutrino flavor amplitudes inside the Sun by taking
into account the roles of the CP-violating phases of the neutrino mixing matrix.
We calculate
the generalized \emph{Parke} formula describing the electron
neutrino survival probability
and the electron-to-sterile neutrino transition
probability. Furthermore, we validate our
analytical calculations through a numerical solution of the
neutrino evolution equation in the case of four-neutrino mixing and we illustrate numerically the
effects on the survival and transition probabilities
of the three CP-violating phases.

This paper is organized as follows.
We review
the general framework of neutrino flavor evolution in Sec.~\ref{general_framework}
and we present the
analytical expressions for the neutrino oscillation
probabilities in Sec.~\ref{sec:parke}.
In Sec.~\ref{sec:numeric} a numerical validation of the
analytical results and an illustration of the effects of the CP-violating
phases is presented in the simplest case of four-neutrino
mixing. Finally, we conclude in Sec.~\ref{sec:conclusion}.
At the end of the paper there are three appendices on the
analytical derivation of the non-adiabatic crossing probability,
on the explicit parametrization of the mixing matrix
and on the density matrix method.

\section{General Framework}
\label{general_framework}

Following the notation of Ref.~\cite{solarste}, in this Section we
shall give a brief review on the general framework of neutrino
flavor evolution with three active and $N_{s}$ sterile neutrinos.
The flavor eigenstates of active neutrinos and sterile neutrinos can
be written as
\begin{equation}
  \left|{\nu^{}_\alpha}\right\rangle = \sum^{N}_{i=1} U^*_{\alpha i}
  \left| {\nu^{}_i} \right\rangle \; ,
\end{equation}
where $N=3+N_{s}$, $\alpha$ runs over $e$, $\mu$ and $\tau$ for
three active neutrinos and $s_{1},\ldots,s_{N_{s}}$ for $N^{}_s$
sterile neutrinos, $\nu^{}_i$ is one of the $N$ mass eigenstates
with mass $m_{i}$, and $U^{}_{\alpha i}$ stands for an element of
the $(3+ N^{}_s)\times (3+ N^{}_s)$ neutrino mixing matrix. In the
general case, solar neutrinos are described by the state
\begin{equation}
  | \nu(x) \rangle = \sum_{\alpha=e,\mu,\tau,s_{1},\ldots,s_{N_{s}}}
  \psi_{\alpha}(x) | \nu_{\alpha} \rangle
  \quad,
\end{equation}
where $x$ is the distance from the production point during the
propagation with the initial condition $\psi_{\alpha}(0) =
\delta_{\alpha e}$ and the normalization $ \sum_{\alpha}
|\psi_{\alpha}(x)|^2 = 1 $. The Mikheev-Smirnov-Wolfenstein (MSW)
equation\cite{Wolfenstein:1978ue,Mikheev:1985gs} describing the
evolution of the flavor transition amplitudes $\psi_{\alpha}(x)$ is
given by (see Ref.~\cite{Giunti-Kim-2007})
\begin{equation}
  i \ddx \Psi = \mathcal{H}_{\rm F}\Psi
  =
  \left( U \mathcal{M}
  U^{\dagger} + \mathcal{V} \right) \Psi
  \quad, \label{eq:MSW_eq}
\end{equation}
with
\begin{align}
  \Psi = \null & \null \left(
    \psi_{e},\psi_{\mu},\psi_{\tau},\psi_{s_{1}},\ldots,\psi_{s_{N_{s}}}
  \right)^T
  \quad, \\
  \mathcal{M} = \null & \null \mathrm{diag} \! \left( 0,
    \frac{\Delta{m}^2_{21}}{2 E}, \frac{\Delta{m}^2_{31}}{2 E}, \frac{\Delta{m}^2_{41}}{2 E},
    \ldots,\frac{\Delta{m}^2_{N1}}{2 E}\right)
  \quad, \\
  \mathcal{V} = \null & \null \mathrm{diag} \! \left(
    \Vcc+\Vnc, \Vnc, \Vnc,
    0, \ldots ,0 \right)
  \quad,
  \label{eq:Vmat}
\end{align}
where $E$ is the neutrino energy and $ \Delta{m}^2_{kj} = m_{k}^2 - m_{j}^2\,$. The
charged-current and neutral-current matter potentials are defined as
\begin{equation}
  \Vcc=\sqrt{2}G_{\mathrm{F}}N_{e} \simeq 7.63 \times 10^{-14}
  \, \frac{ N_{e} }{ N_{\mathrm{A}} \, \mathrm{cm}^{-3} } \, \mathrm{eV}
  \quad, \qquad \Vnc=-\half\sqrt{2}G_{\mathrm{F}}N_{n}
  \quad,
\end{equation}
where $G_{\mathrm{F}}$ is the Fermi constant, $N_{e}$ is the
electron number density, $N_{n}$ is the neutron number density, and
$N_{\mathrm{A}}$ is the Avogadro's number. By using the definition
of electron fraction $Y_{e} = N_{e}/( N_{e}+N_{n})$, we have
\begin{equation}
  N_{e} = \frac{ \rho }{ \mathrm{g} } \, N_{\mathrm{A}} Y_{e}
  \quad, \qquad
  N_{n} = \frac{ \rho }{ \mathrm{g} } \, N_{\mathrm{A}} \left( 1 - Y_{e}
  \right)
  \quad,
\end{equation}
in an electro-neutral medium, with $\rho$ being the mass density.
Thus, we have the following relation between the two matter potentials:
\begin{equation}
  \Vnc = \Rnc \Vcc \quad, \quad \mathrm{with}
  \quad \Rnc = - \frac{ 1 - Y_{e} }{ 2 Y_{e} }
  \quad. \label{eq:Rnc}
\end{equation}
Next, we can introduce the vacuum mass basis as
\begin{equation}
  \label{eq:psiVtoF}
  \PsiV = \left(\psiV_{1}
    ,\ldots,\psiV_{N} \right)^T = U^{\dagger} \Psi
  \quad,
\end{equation}
which satisfies the following evolution equation
\begin{equation}
  \label{eq:psiV_evol}
  i \ddx \PsiV = \left( \mathcal{M} +
    U^{\dagger} \mathcal{V} U \right) \PsiV
  \quad.
\end{equation}
Then we can decouple the flavor transitions generated by
$\Delta{m}^2_{21} $ from those generated by the larger mass-squared
differences, according to the following hierarchy\footnote{ The
different case of active-sterile neutrino mixing with a much smaller
mass-squared difference in solar neutrino oscillations, has been
studied in Ref.~\cite{smirnov}.}
\begin{equation}
  V_{\mathrm{CC}} \sim |V_{\mathrm{NC}}| \sim  \frac{\Delta{m}^2_{21}}{2 E}
  \ll \frac{ |\Delta{m}^2_{k1}| }{2 E} \quad \mathrm{for} \quad k \geq 3
  \label{eq:approx}
  \quad,
\end{equation}
for the solar matter density. Therefore, the N-component evolution
equation Eq.~(\ref{eq:psiV_evol}) can be truncated to
\begin{align}
  i \ddx
  \pVm
  &=
\begin{pmatrix}
    \sum_{\alpha} |U_{\alpha 1}|^2 \mathcal{V}_{\alpha} & \sum_{\alpha} U_{\alpha 1}^{*} U_{\alpha 2} \mathcal{V}_{\alpha}
    \\
    \sum_{\alpha} U_{\alpha 2}^{*} U_{\alpha 1} \mathcal{V}_{\alpha} & \Delta m^2_{21}/(2E) + \sum_{\alpha} |U_{\alpha2}|^2 \mathcal{V}_{\alpha}
  \end{pmatrix}
  \pVm
  \quad, \label{eq:sector12_evol}
\end{align}
and
\begin{align}
  \psiV_{k}(x) &\simeq \psiV_{k}(0) \, \exp\!\left(
    - i \, \frac{ \Delta{m}^2_{k1} x }{ 2 E } \right) \quad, \quad
  \mathrm{for} \quad k \geq 3
  \quad. \label{eq:3more}
\end{align}
By subtracting a diagonal term
\begin{equation}
  \frac{\Delta{m}^2_{21}}{4E}
  +
  \half \sum_{\alpha} \left( |U_{\alpha1}|^2 + |U_{\alpha2}|^2 \right) \mathcal{V}_{\alpha}
  \quad,
\end{equation}
which generates an irrelevant common phase, the evolution equation
in Eq.~(\ref{eq:sector12_evol}) can be written as
\begin{align}
  \label{eq:evol_eq}
  i \ddx \PsiV_2 &= \HV_2 \PsiV_2
  \quad,
\end{align}
with $\PsiV_2=(\psiV_{1}, \psiV_{2})^{\rm T}$ and
\begin{align}
 \HV_2=
  \begin{pmatrix}
    - \delta + V \cos 2 \xi & V \sin 2 \xi e^{i\varphi} \\
    V \sin 2 \xi e^{-i\varphi} & \delta - V \cos 2 \xi
  \end{pmatrix}
  \quad, \label{eq:HV}
\end{align}
where the variables $\delta$, $V$ and $\xi$ are defined as
\begin{align}
  \delta &= \frac{\Delta m^2_{12}}{4E}
  \quad, \\
  V \cos 2 \xi &= \sum_{\alpha} \left( |U_{\alpha1}|^2 - |U_{\alpha2}|^2 \right) \mathcal{V}_{\alpha}
  = \half \Vcc X
  \quad, \\
  V \sin 2 \xi e^{i\varphi} &= \sum_{\alpha} U_{\alpha1}^{*} U_{\alpha2} \mathcal{V}_{\alpha}
  = \half \Vcc Y = \half \Vcc |Y| e^{i\varphi}
  \quad,
\end{align}
with
\begin{align}
  X&= |U_{e1}|^2 - |U_{e2}|^2 + \Rnc \sum_{\alpha=e,\mu,\tau}
  \left( |U_{\alpha1}|^2 - |U_{\alpha2}|^2 \right)
  \nonumber{} \\
  &= |U_{e1}|^2 - |U_{e2}|^2 - \Rnc \sum_{i=1}^{N_{s}}\left(
    |U_{s_{i}1}|^2 - |U_{s_{i}2}|^2 \right)\label{eq:X}
  \quad, \\
  Y&= 2 \left( U_{e1}^{*} U_{e2} + \Rnc
    \sum_{\alpha=e,\mu,\tau} U_{\alpha1}^{*} U_{\alpha2} \right)
  \nonumber{} \\
  &= 2 \left( U_{e1}^{*} U_{e2} - \Rnc \sum_{i=1}^{N_{s}}
    U_{s_{i}1}^{*} U_{s_{i}2} \right)
  \quad. \label{eq:Y}
\end{align}
Therefore, we can obtain the full expressions for $\varphi$,
$\xi$ and $V$ as
\begin{align}
  \varphi &= \arg(Y)
  \quad, \label{eq:varphi} \\
  \tan 2 \xi & = \frac{|Y|}{X}
  \quad, \label{eq:xi} \\
  V &= \half \Vcc \sqrt{ X^2 + |Y|^2 }
  \quad. \label{eq:V}
\end{align}
To be more explicit, we can rewrite the Hamiltonian in Eq.
(\ref{eq:HV}) in a compact form:
\begin{equation}
  \label{eq:HV_mat}
  \HV_2 = \mathcal{M}_2 +
  U^{\dagger}_2 \mathcal{V}_2 U_2
  \quad,
\end{equation}
with $\mathcal{M}_2 = \mathrm{diag}(-\delta,\delta)$, $\mathcal{V}_2
= \mathrm{diag}(V,-V)$ and
\begin{align}
  U_2 &= W_2(\xi,\varphi) \equiv
  \begin{pmatrix}
    \cos\xi & \sin\xi e^{i \varphi} \\
    -\sin\xi e^{-i \varphi} & \cos\xi
  \end{pmatrix}
  \quad,
\end{align}
where $W_2(\xi,\varphi)$ is the complex rotation matrix which can be
generated by a real rotation matrix $R_2(\xi)$ and a diagonal phase
matrix $D_2(\varphi)$ (see Ref.~\cite{Giunti-Kim-2007}),
\begin{align}
  W_2(\xi,\varphi) &= D^{\dagger}_2(\varphi) R_2(\xi) D_2(\varphi)
  \quad,\label{eq:W}
  \\
  R_2(\xi) &\equiv
  \begin{pmatrix}
    \cos\xi & \sin\xi
    \\
    -\sin\xi & \cos\xi
  \end{pmatrix}
  \quad, \label{eq:R} \\
  D_2(\varphi) &\equiv \mathrm{diag}(1,e^{i \varphi})\quad.
  \label{eq:D2}
\end{align}
Using the definition of these effective parameters, we have obtained
an evolution equation analogous to that in the two-neutrino mixing
scheme. But one should keep in mind that the effective mixing angle
$\xi$ and phase $\varphi$ are not simple mixing parameters given by
a specific parametrization but medium-dependent parameters which are
constant only when the electron fraction $Y_e$ remains unchanged along
the neutrino propagation path.

To solve the evolution equation in Eq.~(\ref{eq:evol_eq}), we can
first diagonalize the Hamiltonian in Eq.~(\ref{eq:HV_mat}) with a
complex rotation
\begin{align}
  \PsiV_2 &= W_2(\omega,\varphi) \PsiM_2
  \quad, \label{eq:psiMtoV}
\end{align}
where
\begin{align}
  \tan 2\omega &= \frac{ V \sin2\xi }{ \delta - V \cos2\xi }
  \quad, \label{eq:omega}
\end{align}
$\varphi$ is the complex phase defined in Eq.~(\ref{eq:varphi}), and
$\PsiM_2$ is the amplitude vector in the effective mass basis in
matter. Then the evolution equation becomes
\begin{align}
  i \ddx \PsiM_2 &= \HM_2 \PsiM_2
  \quad, \label{eq:evol_eqM}
\end{align}
where the Hamiltonian can be decomposed into the adiabatic (ad) and
non-adiabatic (na) parts,
\begin{align}
  \HM_2 &= \HMad + \HMna
  \nonumber{} \\
  & \equiv
  \begin{pmatrix}
    -\dM & 0
    \\
    0 & \dM
  \end{pmatrix}
  +
  \begin{pmatrix}
    - \df \sin^2\omega & ( \half \df \sin2\omega - i \dw ) e^{i \varphi}
    \\
    ( \half \df \sin2\omega + i \dw ) e^{-i \varphi} & \df \sin^2\omega
  \end{pmatrix}
  \quad, \label{eq:HM}
\end{align}
with
\begin{align}
  \dM &= \sqrt{ (\delta - V \cos2\xi)^2 + (V \sin2\xi)^2 }
  \nonumber{}\\
  &= \sqrt{ (V - \delta \cos2\xi)^2 + (\delta \sin2\xi)^2 }
  \quad, \label{eq:dM}
\end{align}
\begin{equation}
  \df \equiv \frac{\mathrm{d} \varphi}{\mathrm{d} x} \quad , \quad
  \dw \equiv \frac{\mathrm{d} \omega}{\mathrm{d} x}
  \quad.
\end{equation}
Finally, we can arrive at a formal solution of Eq.
(\ref{eq:evol_eqM}),
\begin{align}
  \left\{
  \begin{array}{ll}
    \overline
    { |\psiM_1(x_d)|^2 } &= |\psiM_1(0)|^2 \left( 1 - P_{12} \right) + |\psiM_2(0)|^2 P_{12}
    \\
    \overline
    { |\psiM_2(x_d)|^2 } &= |\psiM_1(0)|^2 P_{12} + |\psiM_2(0)|^2 \left( 1 - P_{12} \right)
  \end{array} \right.  \label{eq:P12def}
\quad\,,
\end{align}
for the averaged amplitudes of solar neutrino evolution, where $x_d$
is the coordinate of the detector on the Earth, $P_{12}$ is the
level-crossing probability between two effective mass eigenstates
$\psiM_1$, $\psiM_2$ during their propagation inside the Sun. By
definition, $P_{12}$ is zero when the off-diagonal terms in Eq.
(\ref{eq:HM}) are vanishing, which is defined as the adiabatic
approximation.

Before finishing this Section, we want to point out that, as an
improvement to the results in Ref.\cite{solarste}, in this paper we
consider the general case in which the effective phase $\varphi$
cannot be absorbed by a simple rephasing transformation when the
electron fraction $Y_{e}$ is not constant along the neutrino path.
In this case, the CP-violating phases in the mixing matrix may
influence the neutrino flavor evolution inside the Sun through the
effective phase $\varphi$, which is the complex argument of $Y$
Eq.~(\ref{eq:Y}). Moreover, since also the module of $Y$ depends on
the CP-violating phases in $U_{\alpha1}$ and $U_{\alpha2}$, these
phases can affect also on the behavior of $\xi$, $V$ and $\omega$.
In the next Section we derive the analytic expression for the
average oscillation probabilities of solar neutrinos taking into
account the effects of the CP-violating phases in the mixing matrix.

\section{Generalized \emph{Parke} Formula}
\label{sec:parke}

In this Section, we derive the neutrino
oscillation probabilities based on the framework presented in the
previous Section. Due to the energy resolution of the detector and
the uncertainty of the production region, the interference terms
between the massive neutrinos are not measurable \cite{washout}. As
a result, we obtain the averaged oscillation probabilities
\begin{align}
    \averP{S}
  =  &  \overline{ |\psi_{\beta}(x_d)|^2 }
  = \overline{ \left| \sum_{k=1}^{N} U_{\beta k}
      \psiV_{k}(x_d) \right|^2 }
  \nonumber{}\\
  =& \sum_{k=1}^N |U_{\beta k}|^2 \overline{ |\psiV_k(x_d)|^2 }
  \nonumber{}\\
  =& \sum_{k=1}^2 |U_{\beta k}|^2 \overline{ |\psiM_k(x_d)|^2 } + \sum_{k=3}^N |U_{\beta k}|^2 |\psiV_k(0)|^2
  \quad,\label{eq:averP_S_PsiV}
\end{align}
where the matter effects in the detector are neglected. According to
Eq.~(\ref{eq:3more}), $\psiV_k$ with ($k \geq 3$) is decoupled from
other amplitudes inside the Sun and evolves independently. All we
need is to solve the evolution equation in the truncated 1-2 sector,
given in Eq.~(\ref{eq:evol_eq}) in the vacuum basis or in
Eq.~(\ref{eq:evol_eqM}) in the effective mass basis in matter. From
the initial conditions for the flavor amplitudes $ \psi_\beta(0) =
\delta_{\beta e} $, one obtain that in the vacuum basis $ \psiV_k(0)
= U_{ek}^* $.

Writing $U_{\beta 1}$ and $U_{\beta 2}$ as
\begin{equation}
  \left\{\begin{array}{ll}
      U_{\beta 1}&=\cos\theta_\beta \, \cos\chi_\beta \, e^{i\,\phi_{\beta 1}}\\
      U_{\beta 2}&=\sin\theta_\beta \, \cos\chi_\beta \, e^{i\,\phi_{\beta 2}}
    \end{array}\right.\quad
  \text{with} \quad \cos^2\chi_{\beta} = |U_{\beta 1}|^2 + |U_{\beta 2}|^2
  \quad, \label{eq:U_theta_chi}
\end{equation}
the initial conditions in the
effective mass basis in matter are
\begin{align}
  \left\{
     \begin{array}{ll}
       \psiM_1(0) &= \cos\chi_e ( \cos\omega^0 \cos\theta_e e^{-i \phi_{e1}} - \sin\omega^0 \sin\theta_e e^{-i (\phi_{e2} - \varphi^0)} )
       \\
       \psiM_2(0) &= \cos\chi_e ( \sin\omega^0 \cos\theta_e e^{-i (\varphi^0 + \phi_{e1}) } + \cos\omega^0 \sin\theta_e e^{-i \phi_{e2}} )
     \end{array}
  \right.\quad,
\end{align}
where $\omega^0$ and $\varphi^0$ are the rotation parameters [see
the definition in Eq.~(\ref{eq:psiMtoV})] between the vacuum mass
basis and effective mass basis at the production point.

Using the relations in Eq.~(\ref{eq:P12def}), we obtain the averaged
solar neutrino oscillation probabilities
\begin{align}
  \averP{S} &= \cos^2\chi_e \cos^2\chi_{\beta} \averP{(2\nu)} + \sum_{k=3}^{N} |U_{ek}|^2 |U_{\beta k}|^2
  \quad, \label{eq:averP_S}
\end{align}
with
\begin{align}
  \averP{(2\nu)} &\equiv  \half + ( \half -  P_{12} )\cos2\theta_{\beta}
    [ \cos2\theta_e \cos2\omega^0 - \cos\Phi^0 \sin2\theta_e \sin2\omega^0 ]
  \quad, \label{eq:averPS2f}
\end{align}
where
\begin{equation}
\Phi^0 = \phi_{e1} - \phi_{e2} + \varphi^0
\,.
\label{eq:Phi0}
\end{equation}

One can check that the survival probability reduces to the
well-known \emph{Parke} formula \cite{parke} in the limit of
two-neutrino mixing, in which $\cos^2\chi_e = \cos^2\chi_{\beta} =
\cos\Phi^0 = 1$ and $ [ \cos2\theta_e \cos2\omega^0 - \cos\Phi^0
\sin2\theta_e \sin2\omega^0 ] \to \cos2\theta_e^0 $, where $
\theta_e^0 = \theta_e + \omega^0$ is the effective mixing angle
between the flavor basis and the effective mass basis in matter at
the production point.

The oscillation probabilities reduce to the ones in Ref.~\cite{solarste}
in the case of CP invariance, in which $\cos\Phi^0=1$.
Note, however,
that in the case of CP violation
the expression (\ref{eq:averP_S}) for the oscillation probabilities
do not reduce to the corresponding
Eq.~(61) of Ref.~\cite{solarste}
even if the electron fraction $Y_e$ is constant along the neutrino path.
In this case,
as explained in Ref.~\cite{solarste} one can eliminate the phase $\varphi$
by an appropriate rephasing of the mixing matrix.
In fact,
since the MSW evolution equation (\ref{eq:MSW_eq}) is invariant under the phase transformation
\begin{equation}
U_{\alpha k} \to e^{i \varphi_\alpha} U_{\alpha k} e^{i \varphi_k}
\,,
\label{eq:phase}
\end{equation}
which transforms $ \varphi \to \varphi + \varphi_2 - \varphi_1$, one
can eliminate $\varphi$ by choosing $\varphi_1 - \varphi_2 =
\varphi$. However, in this case there is no remaining phase freedom
to eliminate both $\phi_{e1}$ and $\phi_{e2}$, contrary to what has
been incorrectly stated in Ref.~\cite{solarste}. Indeed, since
$\phi_{e1} - \phi_{e2} + \varphi$ is invariant under the phase
transformation (\ref{eq:phase}) (as well as $X$ and $|Y|$, and hence
$\averP{S}$), it is clear that it cannot be eliminated. Hence, the
oscillation probabilities in Eq.~(\ref{eq:averP_S}) represents the
improvement of the corresponding oscillation probabilities in
Eq.~(61) of Ref.~\cite{solarste} which takes into account in a
proper way the effect of the CP-violating phases in the mixing
matrix \footnote{Some effects of the CP-violating phases in solar
neutrino active-sterile oscillations have been discussed earlier in
Refs.~\cite{Schwetz4,palazzo1,palazzo2}.}.

The CP-violating phases can
contribute to the oscillation probabilities in several different
aspects. It is obvious to identify $\cos\Phi^0$ as a term which gives a direct impact,
but in practice $\cos\Phi^0$ is very close to unity for solar neutrinos,
as shown in the next Section
(see, for example, Fig.~\ref{fig:P2f_CP1}).
However, there can be a significant phase dependence of the oscillation probabilities coming from
$\omega^0$, which depends on the CP-violating phases through $X$ and
$|Y|$ and through the modules of the mixing matrix elements in two
different rows, which depend on the cosines of the CP-violating phases.

When counting the relevant number of CP-violating phases in solar
neutrino oscillations, one should take into account that $\nu_\mu$
and $\nu_\tau$ are indistinguishable and all sterile neutrinos are
indistinguishable. Therefore, if the mixing matrix is written as a
product of complex rotations, the rotation in the
$\nu_\mu$-$\nu_\tau$ sector and all the complex rotations among
sterile neutrinos do not have any effect on the observables in solar
neutrinos. These complex rotations can be eliminated from the
evolution equation (\ref{eq:psiV_evol}) by choosing a
parameterization of the mixing matrix in which they occupy the
left-most positions. Then, the remaining part of the mixing matrix
can be written in terms of two complex rotations among active
neutrinos and three complex rotations between the three active
neutrinos and each sterile neutrino, with a total of $2+3N_{s}$
complex rotations. Of the corresponding phases, there are $2+N_{s}$
Majorana phases which can be factorized in a diagonal matrix on the
right of the mixing matrix (see Section~6.7.3 of
Ref.~\cite{Giunti-Kim-2007}) and have no effect on oscillations.
Therefore, solar neutrino oscillations depend on $2N_{s}$ Dirac
CP-violating phases.
\begin{figure}
\begin{center}
\begin{tabular}{c}
\includegraphics*[bb=50 55 400 296, width=0.7\textwidth]{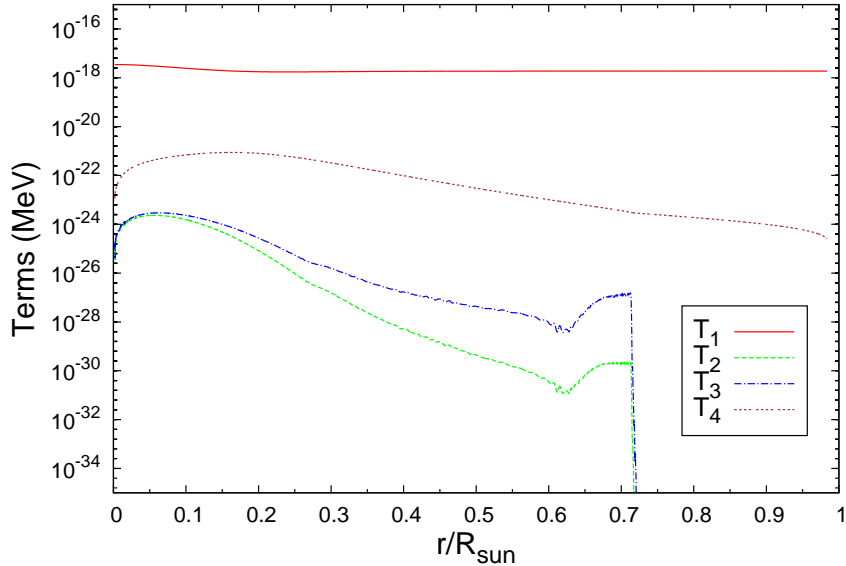}
\end{tabular}
\end{center}
\caption{\label{fig:na_terms}
The magnitudes of four different terms in Eq.~(\ref{eq:HM}):
$T_1=\dM$, $T_2=|\df\sin^2\omega|$, $T_3=|\half
\df \sin2\omega|$ and $T_4=|\dw|$. The mixing parameters are set to
M1 and P1 in Appendix~\ref{sec:paramU} and the matter density
distribution is taken from the BSB2005(OP) Standard Solar Model
\cite{BSB05}.}
\end{figure}

In conclusion of this Section,
let us discuss the problem of calculation of the crossing probability
$P_{12}$. In Appendix~\ref{sec:P_12} we describe two different approximations which allow us to derive
analytical expressions for $P_{12}$. The first method \cite{solarste},
described in Appendix~\ref{sec:const_Ye},
can be applied when the electron fraction $Y_e$ is approximately constant along the neutrino propagation path.
In this case, we can
employ the similarity of the neutrino flavor evolution in
Eq.~(\ref{eq:evol_eqM}) to that of two-neutrino mixing and obtain the
crossing probability with the help of non-perturbative calculations
\cite{pcross}.
The second method, described in Appendix~\ref{sec:general_pert},
can be applied when the non-adiabatic
contribution in the Hamiltonian (\ref{eq:HM}) is much smaller than the adiabatic one.
In this case, one can use the general perturbation theory
\cite{liao,Akhmedov-2004} and calculate the effective crossing
probability to include the non-adiabatic contribution along the
whole path of neutrino propagation inside the Sun.

Both of the above approximations are quite good for the solar
neutrino evolution inside the Sun. As shown in Fig. 1 of Ref.
\cite{solarste}, the electron fraction $Y_{e}$ is almost constant in
the radiative ($0.25\lesssim r \lesssim 0.7$) and convective
($r\gtrsim 0.7$) zones, with the only exception of the core region
with $r\lesssim 0.25$. However, the effects of the variation of
$Y_{e}$ in the core are negligible if the flavor transitions occur
mainly in a resonance located in the radiative or convective zone.
The numerical analysis in Ref. \cite{solarste} validated this
approximation. For the approximation of perturbative expansion, we
show the magnitudes of different terms of Eq.~(\ref{eq:HM}) in
Fig.~\ref{fig:na_terms}, considering, as an example, the values of
the mixing parameters M1 and P1 in Appendix~\ref{sec:paramU} and
using the matter density distribution in the BSB2005(OP) Standard
Solar Model \cite{BSB05}. One can see that the order of magnitude of
the non-adiabatic terms can only reach at most about $0.1\%$ of the
adiabatic term, verifying the accuracy of the perturbative
approximation. In practice, the numerical calculations of $P_{12}$
with the two methods are consistent and both show that $P_{12}$ is
negligibly small. Therefore, in the numerical analysis discussed in
the following Section we neglect the crossing probability $P_{12}$.

\section{Numerical Discussion}
\label{sec:numeric}
\begin{figure}
\begin{center}
\begin{tabular}{c}
\includegraphics*[bb=48 50 402 400, width=0.59\textwidth]{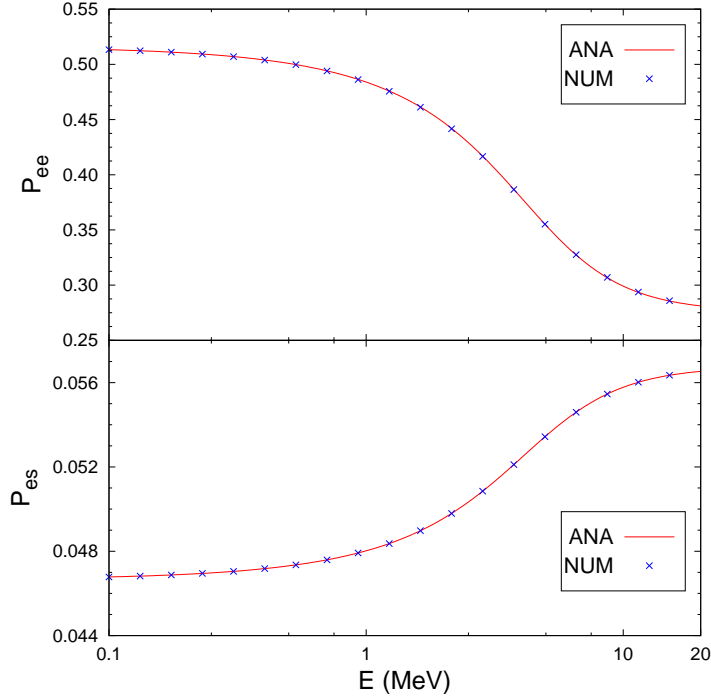}
\end{tabular}
\end{center}
 \caption{\label{fig:Validation}
 Energy spectra of the analytical and numerical evaluations of the solar neutrino electron
 survival (upper panel) and electron-to-sterile transition (lower panel)
 probabilities. All the oscillation parameters are set to M1 and P1
 in Appendix~\ref{sec:paramU}.}
\end{figure}

In this Section, we illustrate the validity of the neutrino
oscillation probabilities in Eq.~(\ref{eq:averP_S}) and the effects
of the CP-violating phases by using a numerical calculation of the
neutrino evolution equation in the scheme of four-neutrino mixing.
The explicit parametrization of the $4\times4$ neutrino mixing
matrix and the values of the oscillation parameters used in the
discussion are presented in Appendix~\ref{sec:paramU}.

As explained in Section~\ref{sec:parke}, in the case $N_{s}=1$ the
observable effects of solar neutrino oscillations depend on two
Dirac CP-violating phases. In fact, we can write the mixing matrix
as\footnote{
This is a standard trick which is used in phenomenological studies
of neutrino oscillations in matter
(see, for example, the three-neutrino mixing discussion in
Section 3.2 of Ref.~\cite{Giunti-Kim-2007}).
It has already been discussed and applied to four-neutrino mixing in
Refs.~\cite{Dooling:1999sg,Giunti:2000wt,palazzo1}.
It allows to eliminate
$W^{23}$
[or
$R^{23}=W(\theta_{23},\eta_{23}=0)$]
from the neutrino evolution equation in all neutrino mixing schemes,
because
$W^{23}$
commutes with the matter potential matrix
$\mathcal{V}$ in Eq.~(\ref{eq:Vmat}).
}
$U=W^{23}U^{\prime}$, where $W^{23}=W(\theta_{23},\eta_{23})$ is
defined in Eq.~(\ref{eq:Wab}) and $U^{\prime}$ is a proper product
of the other rotations, which contains two Dirac CP-violating
phases. Since $W^{23}$ drops out of the evolution equation
(\ref{eq:psiV_evol}), the oscillation probabilities are independent
from $\eta_{23}$ (as well as from $\theta_{23}$). They depend only
on the two Dirac CP-violating phases in $U^{\prime}$. However, in
the following we will discuss the possibility to reveal the effects
of the phases in a future scenario in which the absolute values
$|U_{\alpha4}|$ of the elements of the mixing matrix with $\alpha=e,
\mu, \tau$ have been determined by precision short-baseline neutrino
oscillation experiments. Hence, we adopt the parametrization in
Appendix~\ref{sec:paramU} in which $|U_{e4}|$, $|U_{\mu4}|$ and
$|U_{\tau4}|$ are independent of the phases and determine the mixing
angles $\theta_{14}$, $\theta_{24}$ and $\theta_{34}$ (there is no
way to get such result with $U=W^{23}U^{\prime}$). Hence, although
in the following we consider the three CP-violating phases in the
parametrization in Appendix~\ref{sec:paramU}, one should keep in
mind that the oscillation probabilities depend only on two phases,
which are complicated functions of the three CP-violating phases and
of the mixing angles in the parametrization in
Appendix~\ref{sec:paramU}.

We employ the data of the matter density distribution in the
BSB2005(OP) Standard Solar Model \cite{BSB05} and we consider, for
simplicity, neutrinos produced at the solar center. To obtain the
numerical evolution of solar neutrinos inside the Sun, we use the
fourth-order Runge-Kutta method described in Numerical Recipes
\cite{NR}. Since the unitarity condition is not automatically
guaranteed in a straightforward application of the evolution
equation in Eq.~(\ref{eq:evol_eq}) and can be violated by the errors
of the numerical computation, especially for the evolution in the
crucial resonance region where the amplitudes oscillate rapidly, we
employ the equivalent density matrix formalism (e.g., see Chapter 9
in Ref. \cite{Giunti-Kim-2007}) in which the unitarity condition is
fulfilled by definition. We refer to Appendix~\ref{sec:density} for
a brief introduction on the basics of the density matrix method.

In Fig.~\ref{fig:Validation} we compare the analytical forms (ANA)
of the electron neutrino survival probability $P_{ee} \equiv
\overline{P}^{\mathrm{S}}_{\nu_{e} \to \nu_{e}}$ and the
electron-to-sterile neutrino transition probability $P_{es} \equiv
\overline{P}^{\mathrm{S}}_{\nu_{e} \to \nu_{s}}$ given by
Eq.~(\ref{eq:averP_S}) with the corresponding numerical evaluations
(NUM) of the neutrino flavor transitions. The upper and lower panels
represent the electron survival and electron-to-sterile transition
probabilities, respectively. We illustrate the comparisons with
solid lines and cross points for the analytical and numerical
oscillation probabilities, which show a perfect agreement between
two different calculations of the evolution equation. Numerically,
the accuracy of the analytical calculation is better than $10^{-5}$
and no systematic deviation appears.

Next, we want to illustrate the effects of the CP-violating phases
in solar neutrino active-sterile oscillations. We can observe from
Eq. (\ref{eq:averP_S}) that the oscillation probabilities are only
sensitive to the absolute values of the elements of the neutrino
mixing matrix, except for the explicit contribution of the phases in
$\cos \Phi^0$. However, as discussed in the last Section, the
CP-violating phases determine also the contributions of the modules
of the elements of the mixing matrix, since two distinct rows (i.e.,
the electron and sterile rows) are involved in the oscillation
probabilities\footnote{
See also the discussion in Appendix~D of Ref.~\cite{palazzo1}.
}.
In the specific parametrization of the neutrino
mixing matrix presented in Appendix~\ref{sec:paramU}, the modules of
the matrix elements in the electron row are independent of the
CP-violating phases, but those in other rows are phase-dependent.
Therefore, the effects of the CP-violating phases in the electron
neutrino survival probability arise only in the effective
two-neutrino oscillation probability in Eq. (\ref{eq:averPS2f}) via
the effective mixing parameters $\omega^0$ and $\Phi^0$. On the
other hand, the CP-violating phases manifest themselves in the
electron-to-sterile neutrino transition probability by the phase
dependence in $\omega^0$, $\Phi^0$, $\theta_s$, $\chi_s$. Note that
the constant term in Eq. (\ref{eq:averP_S}) (i.e.,
$\sum_{k=3}^{4}|U_{ek}|^2|U_{sk}|^2$) depends on the variations of
the CP-violating phases via $\chi_s$ in our specific
parametrization.

\begin{figure}
\begin{center}
\begin{tabular}{c}
\includegraphics*[bb=50 48 402 375, width=0.60\textwidth]{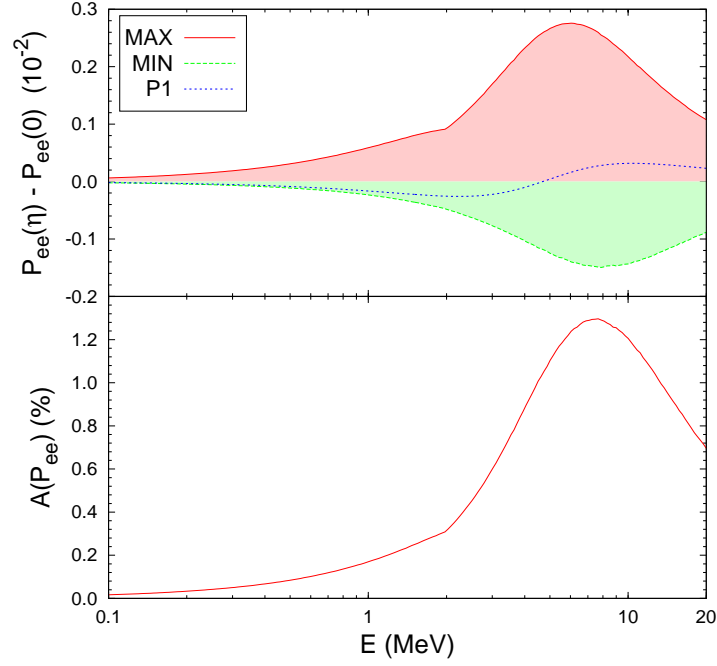}
\end{tabular}
\end{center}
 \caption{\label{fig-En_A1}
 Energy spectra of ${\rm P}_{ee}(\eta)-{\rm P}_{ee}(0)$ (upper panel)
 and $A({\rm P}_{ee})$ (lower panel) for the electron neutrino survival probability.
 The mass and mixing parameters are set to M1 in Appendix~\ref{sec:paramU}
 and the three CP-violating phases are randomly scanned in the full parameter space.
 The short-dashed line in the upper panel corresponds to the set of phases P1  in Appendix~\ref{sec:paramU}.}
\end{figure}
\begin{figure}
\begin{center}
\begin{tabular}{c}
\includegraphics*[bb=50 48 402 375, width=0.60\textwidth]{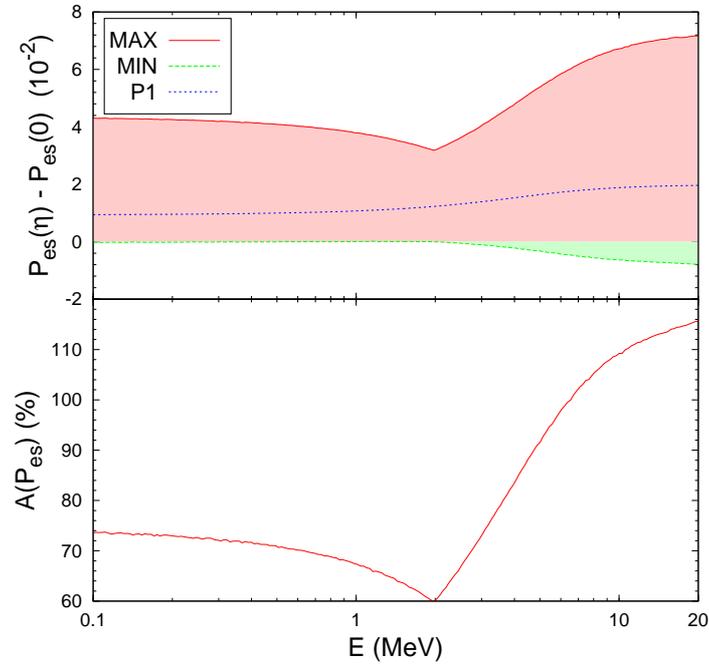}
\end{tabular}
\end{center}
 \caption{\label{fig-En_A2}
 Same as Fig.~\ref{fig-En_A1}, but for the electron-to-sterile neutrino transition probability.}
\end{figure}
\begin{figure}
\begin{center}
\begin{tabular}{c}
\includegraphics*[width=0.6\textwidth]{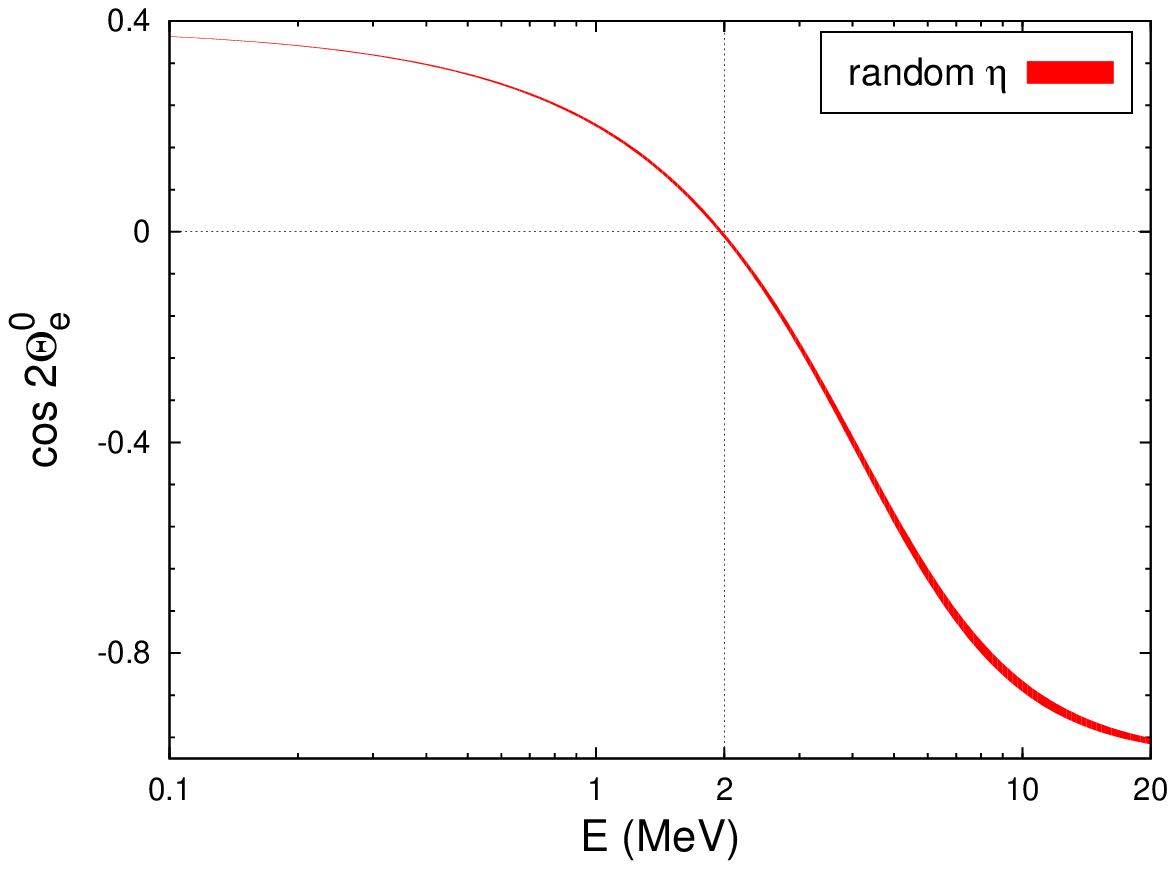}
\end{tabular}
\end{center}
 \caption{\label{fig-En_Theta}
 Energy spectrum of $\cos2\Theta^0_e$ in Eq.~(\ref{c2Te0}).
 The mass and mixing parameters are taken as M1 in Appendix~\ref{sec:paramU}
 and the three CP-violating phases are randomly scanned in the full parameter space. }
\end{figure}

To show the variation of the oscillation probabilities for different
values of the CP-violating phases, we can measure the possible size of the
probability variation as the difference between the maximal (MAX)
and minimal (MIN) values of the probabilities in the full parameter
space of the three CP-violating phases. Therefore, we define the
following asymmetries of the oscillation probabilities:
\begin{align}
  A(Q) = 2 \times \frac{{\rm MAX}[Q] - {\rm MIN}[Q]}{ {\rm MAX}[Q]  + {\rm MIN}[Q]}\,
  \quad,
\end{align}
where $Q$ could be either the survival or transition
probabilities.
Each asymmetry illustrate the possible variation of the corresponding probability
depending on the unknown values of the CP-violating phases
in a future scenario in which the absolute values
$|U_{\alpha4}|$
of the elements of the mixing matrix with $\alpha=e, \mu, \tau$
have been determined by precision short-baseline neutrino oscillation experiments
\cite{whitepaper}.
In the following numerical discussion we consider,
as a realistic example, the values M1 in Eq.~(\ref{M1}) of the mixing angles,
which determine the absolute values of the relevant elements of the mixing matrix
through Eqs.~(\ref{Ue34})--(\ref{Us34}).

In the lower panels of Figs.~\ref{fig-En_A1} and \ref{fig-En_A2} we show the energy
dependence of the asymmetries $A(P_{ee})$ and $A(P_{es})$.
In the upper panels we show the possible range of variation of the probabilities
$P_{ee}$ and $P_{es}$
for all possible values of the phases
$\eta_{14}$,
$\eta_{24}$,
$\eta_{34}$,
with respect to the case
$\eta_{14}=\eta_{24}=\eta_{34}=0$.
The shadowed regions with red and
green colors are generated by scanning the full parameter space of
three CP-violating phases. The two boundary curves stand for the
maximal and minimal values of the differences (which correspond to the
maximal and minimal values of the corresponding probability).
These maximal and minimal values are used in calculating the
corresponding asymmetry $A(P_{ee})$ or $A(P_{es})$ in the
lower panels of Figs.~\ref{fig-En_A1} and \ref{fig-En_A2}.
Note that the boundary curves may
correspond to different values of the CP-violating phases for
different energies.

In the upper panels of Figs.~\ref{fig-En_A1} and \ref{fig-En_A2}
we have also shown the curves corresponding to the values P1 in Eq.~(\ref{P1})
of the CP-violating phases. We can observe
that the variation
induced by these values of the three CP-violating phases is less than $1.2\%$ for the
electron survival probability and can be as large as $100\%$ for the
electron-to-sterile transition probability. This is because the
phase-independent contribution dominates in $P_{ee}$,
whereas
both the phase-independent and phase-dependent contributions are
comparable in $P_{es}$ and both can induce significant
variations in the transition probability. Notice that there is a
kink and a sudden turn at about 2 MeV in the spectra of $A({\rm
P}_{ee})$ and $A({\rm P}_{es})$, respectively, which correspond to the similar behaviour of the
MAX boundary curves in the upper panels.
This property can be
understood with the help of Fig.~\ref{fig-En_Theta},
which shows the energy spectrum of the quantity
\begin{align}
\cos2\Theta_{e}^{0}
\equiv
\cos2\theta_e \cos2\omega^0 - \cos\Phi^0\sin2\theta_e \sin2\omega^0
\label{c2Te0}
\end{align}
in Eq.~(\ref{eq:averPS2f})
with randomly scanned CP-violating phases.
Since $\cos2\Theta^0_e$ changes sign at about 2 MeV,
the values of the CP-violating phases
which maximize the probability have a sudden jump,
which generates a sudden change of the slope of the curve of maximal probability.

\begin{figure}
\begin{center}
\begin{tabular}{c}
\includegraphics*[width=0.9\textwidth]{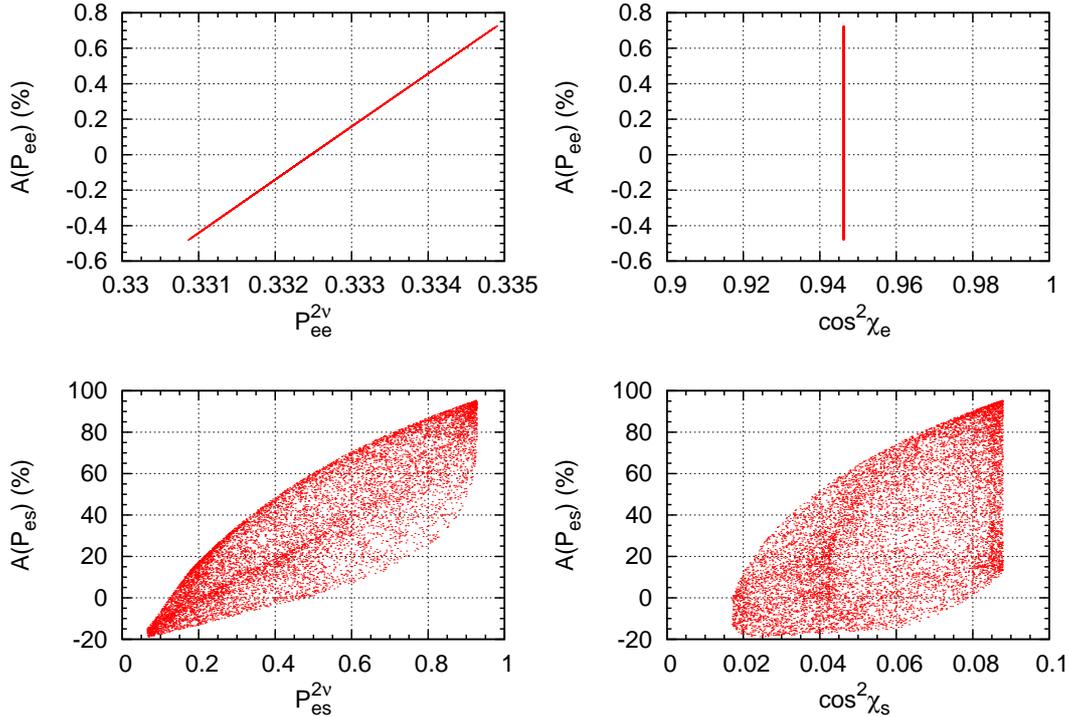}
\end{tabular}
\end{center}
 \caption{\label{fig:P_CP}
 Scatter plots of the asymmetries $A({\rm P}_{ee})$ and $A({\rm P}_{es})$
 with respect to the contributions of the first two mass eigenstates
 (${\rm P}^{2\nu}_{ee}$ or ${\rm P}^{2\nu}_{es}$) and those of all the other mass eigenstates
 ($\cos^2\chi_e$ or $\cos^2\chi_s$). }
\end{figure}
\begin{figure}
\begin{center}
\begin{tabular}{c}
\includegraphics*[width=0.9\textwidth]{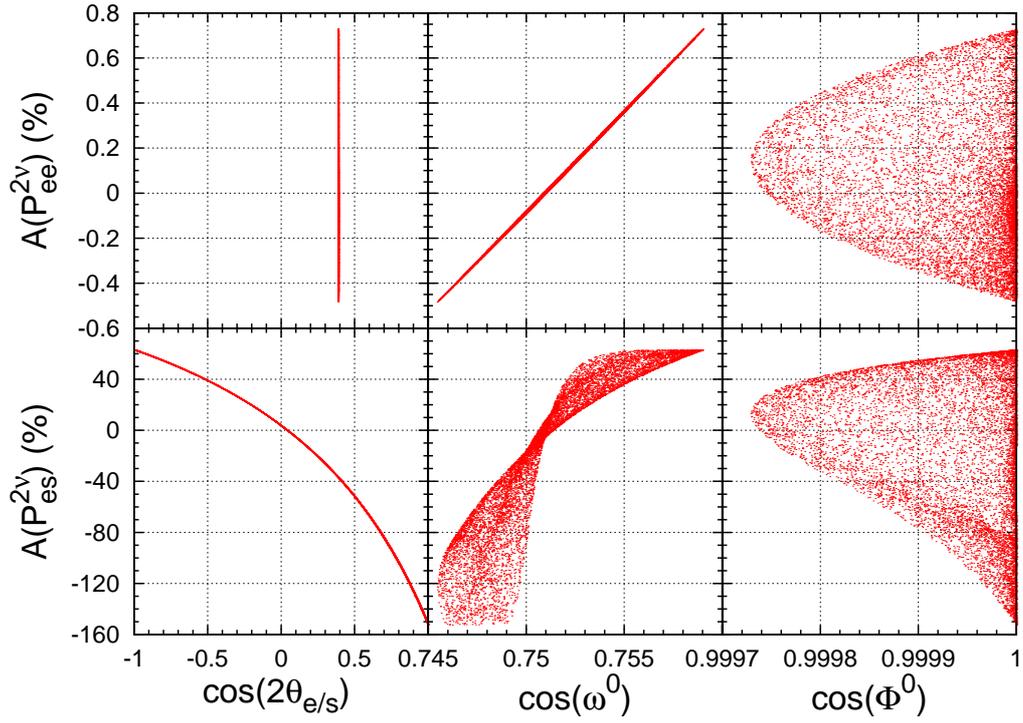}
\end{tabular}
\end{center}
 \caption{\label{fig:P2f_CP1}
 Scatter plots of the asymmetries $A({\rm P}^{2\nu}_{ee})$ and $A({\rm P}^{2\nu}_{es})$
 versus $\cos2\theta_{\rm e/s}$, $\cos\omega^0$ and $\cos\Phi^0$. }
\end{figure}
\begin{figure}
\begin{center}
\begin{tabular}{c}
\includegraphics*[bb=48 50 450 300, width=0.7\textwidth]{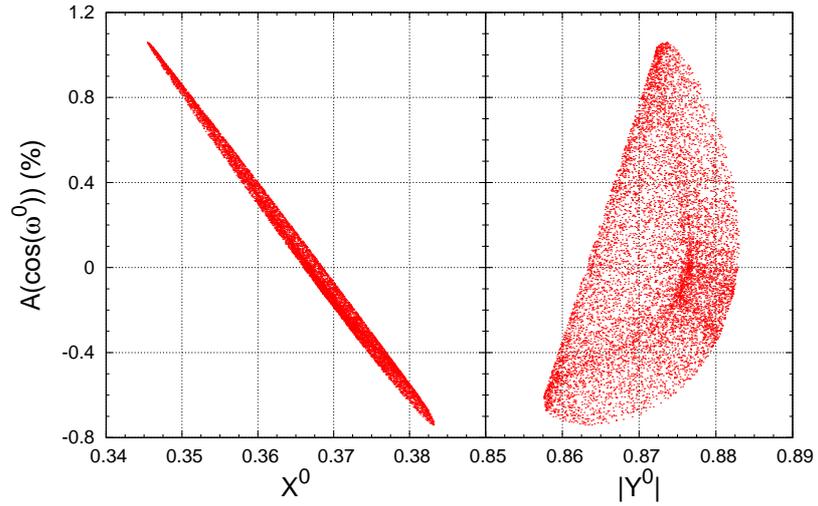}
\end{tabular}
\end{center}
 \caption{\label{fig:P2f_CP2}
 Scatter plots of the asymmetries $A(\cos\omega^0)$
 versus $X^0$ and $|Y^{0}|$. }
\end{figure}

Let us now discuss separately the different contributions to the
probability variation. In Fig.~\ref{fig:P_CP} we show the scatter
plots of $A(P_{ee})$ versus $\cos^2\chi_e$ and ${\rm P}^{2\nu}_{ee}$
and of $A(P_{es})$ versus $\cos^2\chi_s$ and ${\rm P}^{2\nu}_{es}$
obtained with a random generation of the three CP-violating phases
in the entire parameter space. We considered a neutrino energy of 10
MeV and the mixing parameters M1 in Appendix~\ref{sec:paramU}.
Fig.~\ref{fig:P_CP} shows that the effects of the CP-violating
phases show up only in the effective two-neutrino probability ${\rm
P}^{2\nu}_{ee}$ for the survival probability $P_{ee}$, but emerge in
both ${\rm P}^{2\nu}_{es}$ and the suppression factor $\cos^2\chi_s$
induced by the other neutrino states for the transition probability
$P_{es}$. Moreover, we can further study separately the phase
dependence of ${\rm P}^{2\nu}_{ee}$ and ${\rm P}^{2\nu}_{es}$ due to
the phase dependence of $\cos^2\theta_{e/s}$, $\cos\omega^0$ and
$\cos\Phi^0$, which is illustrated in the scatter plots in
Fig.~\ref{fig:P2f_CP1}. In our parametrization of the mixing matrix,
$\cos2\theta_{e}$ is independent of the CP-violating phases, but
$\cos2\theta_{s}$ can reach almost all the possible values with the
varying phase parameters. Therefore, the variation of the
two-neutrino survival probability ${\rm P}^{2\nu}_{ee}$ is dominated
by the changing of $\cos\omega^0$, but the variation of the
two-neutrino transition probability ${\rm P}^{2\nu}_{es}$ comes from
the phase dependence of $\cos2\theta_{s}$ and $\cos\omega^0$. On the
other hand, from the rightmost panels of Fig.~\ref{fig:P2f_CP1},
$\cos\Phi^0$ is very close to unity in both the survival and
transition probabilities. This fact can be explained by considering
Eq.~(\ref{eq:varphi}), where the imaginary part of $Y$ is suppressed
by both the small active-sterile mixing and the small neutral
current contribution ($R_{\rm NC}\simeq -0.2$). Finally, the scatter
plots in Fig.~\ref{fig:P2f_CP2} show the correlations of the
variations of $\cos\omega^0$ with those of $X^0$ and $|Y^0|$. One
can see that the variation of $\cos\omega^0$ is mainly determined by
$X^0$ rather than $|Y^0|$, because of the small phase dependence of
$|Y|$.

In summary, we can conclude that the variation of the survival
probability $P_{ee}$ due to the unknown CP-violating phases in the
mixing matrix is determined mainly by the contribution of
$\omega^{0}$, whereas the transition probability $P_{es}$ is
sensitive to the CP-violating phases via $\chi_s$, $\theta_{s}$ and
$\omega_0$, and the most significant contribution comes from
$\theta_{s}$. We have also shown that the direct phase dependence of
the probabilities through $\cos\Phi^0$ in Eq.~(\ref{eq:averPS2f}) is
negligible because $\cos\Phi^0$ is very close to one in the full
parameter space.

\section{Conclusion}
\label{sec:conclusion}

In this work, we have calculated the analytical solution of the
flavor evolution of solar neutrinos in a general scheme of
3+$N_{s}$ neutrino mixing, without any constraint on the mixing
between the three active and the $N_{s}$ sterile neutrinos. We have improved
the previous study in Ref.~\cite{solarste} by including the possible roles
of the CP-violating phases in the mixing matrix and
we have discussed the effects of these phases
in active-sterile neutrino oscillations.
We derived generalized \emph{Parke}
formulae which are suitable to be used in future
precision measurements of solar neutrino oscillations.

In Section~\ref{sec:numeric} we have presented a
numerical discussion with a realistic example of
the possible phase contribution to the oscillation probabilities
in the case of 3+1 neutrino mixing.
We validated the analytical formulae with a
careful numerical solution of the evolution equation inside the Sun.
We illustrated the effects of the CP-violating phases through an appropriate
asymmetry of the oscillation probabilities.
We have shown that, in our example,
the variations induced by the three unknown
CP-violating phases can reach the level of $1\%$ for the electron
survival probability and may be as large as $100\%$ for the
electron-to-sterile transition probability.
This scenario will be realized when
the absolute values of the elements of the mixing matrix
$|U_{\alpha4}|$ for $\alpha=e, \mu, \tau$
will be measured in precision short-baseline neutrino oscillation
experiments. In this case, it might be possible to observe the
effects of the CP-violating phases in future solar neutrino experiments.

\section*{Acknowledgment}
\label{sec:acknowledgment}
H.W. Long would like to thank Prof.~Peng-Fei ZHANG for his
continuous encouragement and financial support. The work of H.W.
Long is supported in part by the National Natural Science Foundation
of China under Grant No. 11265006. The work of Y. F. Li is supported
in part by the National Natural Science Foundation of China under
Grant No. 11135009.

\newpage
\appendix

\section{Analytical Derivation of $P_{12}$}
\label{sec:P_12}

In this Appendix, we present two methods for the approximate calculation of the crossing
probability $P_{12}$:
the constant $Y_e$ approximation and the approximation of
perturbative expansion.

\subsection{Constant $Y_e$ Approximation}
\label{sec:const_Ye}

Let us consider a case in which $Y_e$ is approximately constant inside the Sun.
Then, $\Rnc$, $\xi$ and $\varphi$ remain approximately unchanged during the neutrino propagation.
Therefore, we can introduce the tilded vacuum mass basis defined by
\begin{align}
  \PsiVt_2 &= D_2(\varphi) \PsiV_2
  \quad, \label{eq:tildeV}
\end{align}
to accommodate the phase $\varphi$ inside the amplitude vector
($D_2(\varphi)$ is defined in Eq.~(\ref{eq:D2})). In
the new basis, the evolution equation in Eq.~(\ref{eq:evol_eq})
becomes
\begin{align}
  i \ddx \PsiVt_2 &= \HVt_2 \PsiVt_2
  \quad, \label{eq:evol_eqVt}
\end{align}
with $\PsiVt_2 = (\tilde{\psi}^{\rm V}_{1}, \tilde{\psi}^{\rm
V}_{2})^{\rm T}$ and
\begin{align}
\HVt_2 =\begin{pmatrix}
    - \delta + V \cos 2 \xi & V \sin 2 \xi \\
    V \sin 2 \xi  & \delta - V \cos 2 \xi
  \end{pmatrix}
  \quad, \label{eq:HVt}
\end{align}
Meanwhile, we can decompose the Hamiltonian into the vacuum and
matter parts as
\begin{align}
  \HVt_2 = \mathcal{M}_2 + \Ut^\dagger \mathcal{V}_2 \Ut
  \quad,
\end{align}
with $\Ut = R_2(\xi)$ [defined in Eq.~(\ref{eq:R})].

As in the discussions in Ref.~\cite{solarste}, we can further
introduce the tilded effective interaction basis $\PsiIt_2$ and
the tilded effective mass basis $\PsiMt_2$ in matter defined by
\begin{align}
  \PsiIt_2 = R_2(\xi) \PsiMt_2\,,
  \quad\quad
  \PsiVt_2 = R_2(\omega) \PsiMt_2
  \,. \label{eq:tildeM}
\end{align}
In this way, we obtain
evolution equations for
$\PsiIt_2$ and $\PsiMt_2$
which have the same form as
those without the CP-violating phases \cite{solarste} if all the
quantities with tildes are replaced by those without tildes. For
instance, in the $\PsiMt_2$ basis the evolution equation can be
written as
\begin{align}
  i \ddx \PsiMt_2 &=
  \begin{pmatrix}
    -\dM & -i\dw \\
    i\dw & \dM
  \end{pmatrix}
  \PsiMt_2
  \quad, \label{eq:evol_PsiMt}
\end{align}
which is just the standard evolution equation of two-neutrino mixing
in matter \cite{Giunti-Kim-2007}. From the similarity we can define
the adiabaticity parameter
\begin{equation}
  \label{eq:gamma}
  \gamma =
  \frac{ \dM }{ | \dw | }
  =
  \frac{ 2 \dM^{\,3} }{ V \delta \sin2\xi | \mathrm{d} \ln N_{e} / \mathrm{d} x | }
  \quad,
\end{equation}
and obtain the crossing probability $P_{12}$ as
\begin{equation}
  P_{12} = \frac{\exp\left( - \frac{\pi}{2} \gamma_{\mathrm{R}} F \right)
  - \exp\left( - \frac{\pi}{2} \gamma_{\mathrm{R}} \frac{F}{\sin^{2}\xi} \right)}
  {1 - \exp\left( - \frac{\pi}{2} \gamma_{\mathrm{R}} \frac{F}{\sin^{2}\xi} \right)} \,
  \theta \left(V_{0}-V_{\mathrm{R}}\right)
  \quad, \label{eq:P12}
\end{equation}
where $\VR = \delta \cos2\xi $ defines the resonance point and
$\gamma_{\mathrm{R}}$ is the adiabaticity parameter at the resonance
with
\begin{equation}
  \gamma_{\mathrm{R}}
  = \frac { 2 \delta \sin^{2}2\xi}
  { \cos2\xi \left|\mathrm{d}\ln N_{e}/\mathrm{d}x\right|_{\mathrm{R}}}
  \quad,
\end{equation}
Finally, the $\theta$ function is used to reduce $P_{12}$ to zero
when the potential at the production point is smaller than that at
the resonance point.
We use $F = 1 - \tan^2 \xi$ for an exponential density
profile, which is a good approximation for the solar neutrinos
\cite{pcross}.

\subsection{Perturbative Expansion}
\label{sec:general_pert}

As discussed in Section 3, the non-adiabatic terms are much smaller
that the adiabatic term and we can treat $\HMna$ in
Eq.~(\ref{eq:HM}) as a perturbation term relative to $\HMad$.
Therefore, we can solve the $S$-matrix defined in the effective mass
basis,
\begin{align}
  \PsiM_2(x_f) = S^{\mathrm{M}}_2(x_f,x_i) \PsiM_2(x_i) \quad, \label{eq:SM}
\end{align}
by using the standard perturbation theory (see Appendix B of Ref.
\cite{Akhmedov-2004}). After a straightforward calculation, we
arrive at the expression of
\begin{align}
  \SMpX &\equiv \SMXad - i \SMXad \intx \SMXada^{-1} \HMna(x) \SMXada \dx
  \nonumber{}\\
  &=\SMXad -i \SMXad
  \begin{pmatrix}
    -A & C \\
    C^{*} & A
  \end{pmatrix}
  \nonumber{} \\
  &=
  \begin{pmatrix}
    (1+iA) e^{i\Delta} & -iC e^{i\Delta} \\
    -iC^* e^{-i\Delta} & (1-iA) e^{-i\Delta}
  \end{pmatrix}
  \quad,\label{eq:SMpX}
\end{align}
where
\begin{align}
  \SMXad &= e^{ -i \intx \HMad(x) \dx }
  \nonumber{} \\
  &=
  \begin{pmatrix}
    e^{i \Delta(x_f,x_i)} & 0 \\
    0 & e^{-i \Delta(x_f,x_i)}
  \end{pmatrix}
  \quad, \\
  A(x_f,x_i) &= \intx \df \sin^2 \omega \dx
  \quad, \label{eq:A} \\
  C(x_f,x_i) &= \intx ( \half \df \sin2\omega - i \dw ) e^{i ( \varphi -2 \Delta(x,x_i) )} \dx
  \quad, \label{eq:C} \\
  \Delta(x_f,x_i) &= \intx \dM \dx
  \quad.
\end{align}
Then, the effective crossing probability $P_{12}$ is given by the probability of
$1 \leftrightarrows 2$
non-adiabatic transitions
\begin{align}
  P_{12} &= \left| \left\{ \SMpert(x_f,0) \right\}_{12} \right|^2
  \nonumber{}\\
  &= |C(x_f,0)|^2 = \left| \int_0^{x_f} ( \half \df \sin2\omega - i \dw ) e^{i ( \varphi -2 \Delta(x,0) )} \dx
  \right|^2\quad.
\end{align}

\section{Explicit Parametrization of $U$}
\label{sec:paramU}

The $4\times4$ neutrino mixing matrix can be parametrized (see Ref.
\cite{Giunti-Kim-2007} for detailed discussion) as an extension of
the standard parametrization \cite{PDG} of three-neutrino mixing:
\begin{align}
  U=W^{34} W^{24} R^{14} R^{23} W^{13} R^{12}
  \quad,\label{eq:Uparam}
\end{align}
where $W^{ab}=W(\theta_{ab},\eta_{ab})$ and
$R^{ab}=W^{ab}(\theta_{ab},0)$ are the complex and real unitary
matrices in the $(a,b)$ plane, where $W(\theta_{ab},\eta_{ab})$ is
defined by
\begin{align}
[W(\theta_{ab},\eta_{ab})]_{rs}=\delta_{rs} &+
(\cos\theta_{ab}-1)(\delta_{ra}\delta_{sa} +
\delta_{rb}\delta_{sb})\nonumber{}\\
&+\sin\theta_{ab}(e^{-i\eta_{ab}}\delta_{ra}\delta_{sb}-e^{i\eta_{ab}}\delta_{rb}\delta_{sa})\quad,
\label{eq:Wab}
\end{align}
with $\theta_{ab}$ and $\eta_{ab}$ being the mixing angles and Dirac
CP phases in the specific plane.

In this parametrization, the explicit expressions for the elements
in the electron and sterile rows of $U$ are given as follows
\begin{align}
  \null & \null
  U_{e1} = c_{12} c_{13} c_{14}
  \,,
  \quad
  \null && \null
  U_{e2} = s_{12} c_{13} c_{14}
  \,,
  \label{Ue12}
  \\
  \null & \null
  U_{e3} = s_{13}e^{-i \eta_{13}} c_{14}
  \,,
  \quad
  \null && \null
  U_{e4} = s_{14}
  \,,
  \label{Ue34}
\end{align}
\begin{align}
  U_{s1} =& - s_{14} c_{12} c_{13} c_{24} c_{34}
  + ( s_{12} c_{23} + s_{13}e^{i\eta_{13}} s_{23} c_{12} ) s_{24} e^{i\eta_{24}} c_{34}
  \nonumber{}\\
  &+ (-s_{12} s_{23} + s_{13}e^{i\eta_{13}} c_{12} c_{23} ) s_{34} e^{i\eta_{34}}
  \,, \label{Us1}\\
  U_{s2} =& - s_{12} s_{14} c_{13} c_{24} c_{34}
  + (-c_{12} c_{23} + s_{12} s_{13}e^{i\eta_{13}} s_{23} ) s_{24} e^{i\eta_{24}} c_{34}
  \nonumber{}\\
  &+ ( s_{23} c_{12} + s_{12} s_{13}e^{i\eta_{13}} c_{23} ) s_{34} e^{i\eta_{34}}
  \,, \label{Us2}\\
  U_{s3} =& -s_{13}e^{-i\eta_{13}} s_{14} c_{24} c_{34}
  - ( s_{34} e^{i\eta_{34}} c_{23} + s_{23} s_{24} e^{i\eta_{24}} c_{34} ) c_{13}
  \,, \label{Us3}\\
  U_{s4} =& c_{14} c_{24} c_{34}
  \,. \label{Us34}
\end{align}
Moreover, we have
\begin{equation}
U_{\mu4} = c_{14} s_{24} e^{-i\eta_{24}}
\,,
\qquad
U_{\tau4} = c_{14} c_{24} s_{34} e^{-i\eta_{34}}
\,.
\label{Um4}
\end{equation}
In our numerical calculations, we consider the following
values of the oscillation parameters:
\begin{align}
\rm{M1:}\quad\left\{\begin{array}{lllll}
\Delta m^2_{12} \simeq 7.54 \times 10^{-5}\,\mathrm{eV}\\
\theta_{12} \simeq 33.6^\circ\\
\theta_{23} \simeq 39.1^\circ\\
\theta_{13}\simeq 9.0^\circ\\
\theta_{14}=\theta_{24}=\theta_{34}=10^\circ\,,
\end{array}\right.\quad
\label{M1}
\end{align}
and
\begin{align}
\rm{P1:}\quad\begin{array}{l}
\eta_{13}=35^\circ\,,\quad\eta_{24}=75^\circ\,,\quad\eta_{34}=115^\circ\quad,
\end{array}
\label{P1}
\end{align}
where the $\Delta m^2_{12}$ and three mixing angles ($\theta_{12}$,
$\theta_{13}$ and $\theta_{23}$), equivalent to the case of
three-neutrino mixing, are taken from the latest global analysis
\cite{fogli}, the active-sterile mixing angles are motivated by the
anomalies of SBL data \cite{LSND,Mini,Reactor,Gallium} and the
phases are chosen non-trivially to reveal the effects of the CP
phases. The assumed values of the active-sterile mixing angles
$\theta_{14}$,
$\theta_{24}$,
$\theta_{34}$
do not significantly affect the values of the oscillation
parameters of active neutrinos extracted from the current data.

\section{Density Matrix Method}
\label{sec:density}

The density matrix formalism is equivalent to the framework of
flavor amplitude evolution in Section 2. We consider that a neutrino
state at the position $x$ is described by the Hermitian
\emph{density matrix operator}
\begin{align}
  \hrx &= \sum_\alpha |\nu_\alpha(x) \rangle W_\alpha \langle \nu_\alpha(x)|
  \quad,\label{eq:hrx}
\end{align}
where $W_\alpha$ is the initial statistical weight of flavor
$\alpha$ (i.e. the probability of the flavor $\alpha$ at $x=0$). One
can choose $W_\alpha=\delta_{\alpha\beta}$ for a neutrino state of
one initial flavor $\beta$. In the flavor basis we can define the
specific \emph{density matrix} as
\begin{align}
  \rF_{\eta \xi} &= \langle \nu_\eta| \hrx |\nu_\xi \rangle = \sum_\alpha W_\alpha \psi_{\alpha\eta}(x) \psi^*_{\alpha \xi}(x)
  \quad.
\end{align}
The evolution equation of the density matrix $\rF$ in the flavor
basis, obtained from the evolution equation in Eq.
(\ref{eq:MSW_eq}), is
\begin{align}
  i \frac{ \mathrm{d} \rF }{ \dx} = \HF \rF - \rF {\HF}
  \quad,\label{eq:rF_evol}
\end{align}
with the initial condition $\rF_{\eta \xi}(0) = W_\eta \delta_{\eta
\xi} $. 
The density matrix in the vacuum mass basis $ \rV= U^\dagger \rF U$
follows the evolution equation
\begin{align}
  i \frac{ \mathrm{d} \rV }{ \dx} = \HV \rV - \rV {\HV}
  \quad,\label{eq:rV_evol}
\end{align}
with $\HV= U^\dagger \HF U$. For solar
neutrino oscillations, using the approximation in
Eq.~(\ref{eq:approx}), we can obtain the reduced evolution equation
\begin{align}
  i \frac{ \mathrm{d} \rV_2 }{ \dx} = \HV_2 \rV_2 - \rV_2 {\HV_2}
  \quad,\label{eq:rV2_evol}
\end{align}
in the $2\times2$ subsystem of $(\psiV_{1}, \psiV_{2})$, where
$\HV_2$ is defined in Eq.~(\ref{eq:evol_eq}) and $\rV_2$ is given by
$ \{\rV_2(x)\}_{ij}=\psiV_{i}(x) {\psiV_j}^*(x)\,$. Since $\HV_2$ is
Hermitian and traceless, from the initial condition $W_\alpha
=\delta_{e\alpha}$ we have
\begin{align}
\mathrm{Tr}[\rV_2(x)] = |\psiV_1|^2 + |\psiV_2|^2 = |U_{e1}|^2 +
|U_{e2}|^2\quad.
\end{align}
Therefore,
the unitarity condition is fulfilled by definition. To be more
explicit, we can rewrite these matrices in the terms of Pauli
matrices with
\begin{align}
  \HV_2 &= - \half \PauliV \cdot \vB
  \quad,\\
  \rV_2 &= \frac{|U_{e1}|^2 + |U_{e2}|^2}{2} \, \mathbf{1} + \half \PauliV \cdot \vS
  \quad.
\end{align}
where
\begin{align}
\PauliV=\sum^3_{a=1}\sigma_{a}\vec{e}_{\rm V}^{\,a}\,,\quad
\vB=\sum^3_{a=1}B^{\rm V}_{a}\vec{e}_{\rm V}^{\,a}\,,\quad
\vS=\sum^3_{a=1}S^{\rm V}_{a}\vec{e}_{\rm V}^{\,a}\quad,
\end{align}
with $\sigma_{a}\,(a=1,2,3)$ being the Pauli matrices and
$(\vec{e}_{\rm V}^{\,1}, \vec{e}_{\rm V}^{\,2}, \vec{e}_{\rm
V}^{\,3})$ being three orthonormal vectors which form the vacuum
mass basis. The components of the vectors $\vB$ and $\vS$ in the
vacuum mass basis are
\begin{align}
  \vB &= \left( -2\,\mRe\{\HV_2\}_{12}, \quad 2\,\mIm\{\HV_2\}_{12}, \quad \{\HV_2\}_{22}-\{\HV_2\}_{11}\, \right)
  \\
  \vS &= \left( 2\,\mRe\{\rV_2\}_{12}, \quad -2\,\mIm\{\rV_2\}_{12}, \quad \{\rV_2\}_{11}-\{\rV_2\}_{22}
  \right)\,.
\end{align}
The evolution equation of the vector $\vS$ is
\begin{align}
  \frac{ \mathrm{d} \vS }{ \dx } &= \vS \times \vB
  \label{eq:graph}
  \quad,
\end{align}
with the initial condition
\begin{align}
  \vS(0) &= \left( 2 \mRe(U^*_{e1}U_{e2}), \quad -2 \mIm(U^*_{e1}U_{e2}), \quad |U_{e1}|^2 - |U_{e2}|^2 \right)
  \quad.
\end{align}
According to Eq.~(\ref{eq:averP_S_PsiV}), the oscillation
probabilities can be written as
\begin{align}
\averP{S} =  \sum_{k=1}^{2} |U_{\beta k}|^2 \{\rV_2(x_f)\}_{kk}
  + \sum_{k=3}^{N} |U_{\beta k}|^2 |U_{e k}|^2
  \quad.\label{eq:averP_S_num}
\end{align}
Using Eqs.~(\ref{eq:graph}) and (\ref{eq:averP_S_num}), we can
employ the fourth-order Runge-Kutta method to perform the numerical
evaluation of the neutrino flavor evolution.



\newpage

\begin{thebibliography}{99}


\bibitem{PDG}
Particle Data Group, (J. Beringer {\it et al.}), Phys. Rev. D {\bf
86}, 010001 (2012).

\bibitem{LSND}
LSND Collaboration, (A. Aguilar {\it et al.}), Phys. Rev. D {\bf
64}, 112007 (2001).

\bibitem{Mini}
MiniBooNE Collaboration, (A.A. Aguilar-Arevalo {\it et al.}), Phys.
Rev. Lett. {\bf 105}, 181801 (2010).

\bibitem{Reactor}
G. Mention {\it et al.}, Phys. Rev. D {\bf  83}, 073006 (2011); P.
Huber, Phys. Rev. C {\bf 84}, 024617 (2011).

\bibitem{Gallium}
C. Giunti and M. Laveder, Phys. Rev. C {\bf 83}, 065504 (2011).

\bibitem{Giunti11}
C. Giunti and M. Laveder, Phys. Rev. D {\bf 84}, 073008 (2011);
Phys. Rev. D {\bf 84}, 093006 (2011).

\bibitem{Giunti12}
C. Giunti {\it et al.}, Phys. Rev. D {\bf 86}, 113014 (2012); Phys.
Rev. D {\bf 87}, 013004 (2013).

\bibitem{Schwetz4}
J. Kopp, M. Maltoni and T. Schwetz, Phys. Rev. Lett. {\bf 107},
091801 (2011);
J.~Kopp, P.~A.~N.~Machado, M.~Maltoni and T.~Schwetz,
  JHEP {\bf 1305}, 050 (2013).

\bibitem{Raffelt}
J. Hamann {\it et al.}, Phys. Rev. Lett. {\bf 105}, 181301 (2010);
E. Giusarma {\it et al.}, Phys. Rev. D {\bf 83}, 115023 (2011); J.
Hamann, JCAP {\bf 1203}, 021 (2012); M. Archidiacono {\it et al.},
Phys.Rev. D {\bf 86}, 065028 (2012).

\bibitem{Ade:2013lta}
Planck Collaboration, (P.~A.~R.~Ade {\it et al.}),
  arXiv:1303.5076 [astro-ph.CO].

\bibitem{Mangano}
G. Mangano and P.D. Serpico, Phys. Lett. B {\bf 701}, 296 (2011); J.
Hamann {\it et al.}, JCAP {\bf 1109}, 034 (2011); T.D. Jacques, L.M.
Krauss and C. Lunardini, Phys. Rev. D {\bf 87}, 083515 (2013).

\bibitem{solarste}
C. Giunti and Y.F. Li, Phys. Rev. D {\bf 80}, 113007 (2009); Prog.
Part. Nucl. Phys. {\bf 64}, 213 (2010).

\bibitem{palazzo1}
A. Palazzo, Phys. Rev. D {\bf 83}, 113013 (2011).

\bibitem{palazzo2}
A. Palazzo, Phys.Rev. D {\bf 85}, 077301 (2012).

\bibitem{ice}
S. Razzaque and A.Y. Smirnov JHEP {\bf 1107}, 084 (2011); V. Barger,
Y. Gao and D. Marfatia, Phys. Rev. D {\bf 85}, 011302 (2012); A.
Esmaili, F. Halzen and O.L.G. Peres, JCAP {\bf 1211}, 041 (2012); R.
Gandhi and P. Ghoshal, Phys. Rev. D {\bf 86}, 037301 (2012).

\bibitem{beta}
A.S. Riis and S. Hannestad, JCAP {\bf 1102}, 011 (2011); J.A.
Formaggio and J. Barrett, Phys.Lett. B {\bf 706}, 68 (2011), A.
Esmaili and O.L.G. Peres, Phys. Rev. D {\bf 85}, 117301 (2012).

\bibitem{double}
Y.F. Li and S.S. Liu, Phys.Lett. B {\bf 706}, 406 (2012);  C. Giunti
and M. Laveder, Phys. Lett. B {\bf 706}, 200 (2011); J. Barry {\it
et al.}, JHEP {\bf1107}, 091 (2011); C. Giunti and M. Laveder, Phys.
Rev. D {\bf 82}, 053005 (2010).

\bibitem{Wolfenstein:1978ue}
L.~Wolfenstein, Phys. Rev. {\bf D17}, 2369 (1978).

\bibitem{Mikheev:1985gs}
S.~P. Mikheev and A.~Y. Smirnov, Sov. J. Nucl. Phys. {\bf 42}, 913
(1985).

\bibitem{whitepaper}
K.N. Abazajian {\it et al.}, (2012) arXiv:1204.5379 [hep-ph].

\bibitem{NU}
S. Antusch, C. Biggio, E. Fernandez-Martinez, M.B. Gavela and J.
Lopez-Pavon, JHEP {\bf 0610}, 084 (2006).

\bibitem{NSI}
T. Ohlsson, Rep. Prog. Phys. {\bf 76}, 044201 (2013).

\bibitem{Giunti-Kim-2007}
C. Giunti and C.W. Kim, {\em Fundamentals of Neutrino Physics and
Astrophysics} (Oxford University Press, Oxford, UK, 2007).

\bibitem{smirnov}
P.C. de~Holanda and A.Y. Smirnov, Phys. Rev. D {\bf 69}, 113002
(2004); Phys. Rev. D {\bf83}, 113011 (2011).

\bibitem{washout}
A.S. Dighe, Q.Y. Liu and A.Y. Smirnov, arXiv:hep-ph/9903329.

\bibitem{parke}
S.J. Parke, Phys. Rev. Lett. {\bf 57}, 1275 (1986).

\bibitem{pcross}
S.T. Petcov, Phys. Lett. B {\bf 200}, 373 (1988); P.I. Krastev and
S.T. Petcov, Phys. Lett. B {\bf 207}, 64 (1988);  {\bf214}, 661(E)
(1988);  S.T. Petcov, Phys. Lett. B {\bf214}, 139 (1988); T.K. Kuo
and J. Pantaleone, Phys. Rev. D {\bf 39}, 1930 (1989).

\bibitem{liao}
P.C. de Holanda, W. Liao and A.Yu. Smirnov, Nucl. Phys. B {\bf 702},
307 (2004); W. Liao, Phys. Rev. D {\bf 77}, 053002 (2008).

\bibitem{Akhmedov-2004}
E.K. Akhmedov, R. Johansson, M. Lindner, T. Ohlsson and T. Schwetz,
JHEP {\bf 0404}, 078 (2004).

\bibitem{BSB05}
J.N. Bahcall, A.M. Serenelli and S. Basu, Astrophys. J. {\bf 621},
L85 (2005).

\bibitem{Dooling:1999sg}
  D.~Dooling, C.~Giunti, K.~Kang and C.~W.~Kim,
  Phys.\ Rev.\ D {\bf 61}, 073011 (2000).

\bibitem{Giunti:2000wt}
  C.~Giunti, M.~C.~Gonzalez-Garcia and C.~Pena-Garay,
  Phys.\ Rev.\ D {\bf 62}, 013005 (2000).

\bibitem{fogli}
G.L. Fogli, E. Lisi, A. Marrone, D. Montanino, A. Palazzo and A.M.
Rotunno, Phys. Rev. D {\bf 86}, 013012 (2012).



\bibitem{NR}
H. William, S. A. Teukolsky, William T. Vetterling, Brian P.
Flannery, {\em {Numerical Recipes in Fortran 77: The Art of
Scientific Computing Second Edition}} (Cambridge University Press).


\end{thebibliography}
\end{document}